\documentclass[structabstract]{aa}  
\usepackage{multirow} 
\usepackage{graphicx}
\usepackage{color}
\usepackage{natbib}
\bibpunct{(}{)}{;}{a}{}{,} %
\usepackage{txfonts}
\begin{document}

   \title{HD 85567: A Herbig B[e] star or an interacting B[e] binary?\thanks{Based on observations conducted at the European Southern Observatory, Paranal, Chile, which were obtained via the program 089.C-0220}}
\subtitle{Resolving HD 85567's circumstellar environment with the VLTI and AMBER}

   \author{H.E. Wheelwright
     \inst{1}
     \and {G. Weigelt}\inst{1}
     \and {A. Caratti o Garatti}\inst{1}
          \and {R. Garcia Lopez}\inst{1}
              }

   \institute{Max-Planck-Institut f\"{u}r Radioastronomie, Auf dem H\"{u}gel 69,
53121 Bonn, Germany\\\email{hwheelwright@mpifr-bonn.mpg.de}
}

   \date{Received month dd, yyyy; accepted Month dd, yyyy}

\abstract
{\object{HD 85567} is an enigmatic object exhibiting the B[e] phenomenon, i.e. an infrared excess and forbidden emission lines in the optical. The object's evolutionary status is uncertain and there are conflicting claims that it is either a young stellar object or an evolved, interacting binary.}
{To elucidate the reason for the B[e] behaviour of HD 85567, we have observed it with the VLTI and AMBER.}
{Our observations were conducted in the $K-$band with moderate spectral resolution (R$\sim$1500, i.e. $200~\mathrm{km\,s^{-1}}$). The spectrum of HD 85567 exhibits Br$\gamma$ and CO overtone bandhead emission. The interferometric data obtained consist of spectrally dispersed visibilities, closure phases and differential phases across these spectral features and the $K-$band continuum.}
{The closure phase observations do not reveal evidence of asymmetry. The apparent size of HD 85567 in the $K-$band was determined by fitting the visibilities with a ring model. The best fitting radius, $0.8\pm0.3$~AU , is relatively small making HD 85567 undersized in comparison to the size-luminosity relationship based on YSOs of low and intermediate luminosity. This has previously been found to be the case for luminous YSOs, and it has been proposed that this is due to the presence of an optically thick gaseous disc. We demonstrate that the differential phase observations over the CO bandhead emission are indeed consistent with the presence of a compact ($\sim$1 AU) gaseous disc interior to the dust sublimation radius.}
{The observations reveal no sign of binarity. However, the data do indicate the presence of a gaseous disc interior to the dust sublimation radius. We conclude that the data are consistent with the hypothesis that HD 85567 is a YSO with an optically thick gaseous disc within a larger dust disc that is being photo-evaporated from the outer edge.}

\titlerunning{HD 85567 observed with VLTI/AMBER}
\authorrunning{H.E. Wheelwright et al.}
   \maketitle
%

\section{Introduction}

The formation and early evolution of massive stars is difficult to
study and, as a result, still not fully understood. This is partly
because sites of massive star formation are typically situated at
greater distances than nearby sites of low mass star formation. In
addition, massive stars form rapidly, deep within their natal
clouds. These factors make detailed study of the small scale
environment of young massive stars challenging. Consequently, our
understanding of how the star formation process depends on mass is
incomplete. To address this issue, it is important to characterise the
circumstellar environment of massive young stars and contrast this
to the case of low mass star formation.

\smallskip

Most studies on the comparison between low and high mass star
formation have focused on Herbig Ae/Be stars. These objects are
pre-main-sequence objects identified by the presence of an infrared
excess and emission lines and have a mass of $\sim2-15$~$M_{\odot}$
\citep{Herbig1960,The1994}. Herbig Ae/Be (HAe/Be) stars span the
transition from low to high stellar masses. Since they are optically
visible and relatively luminous, HAe/Be stars offer an opportunity to
study the circumstellar geometry of young stellar objects (YSOs) at
intermediate and high luminosities. As a result, there have been many
studies of the circumstellar environment of HAe/Be stars \citep[see
  e.g.][]{Meeus2001, Natta2001, Vink2002, Millan-Gabet2001,
  Eisner2004,Acke2005,
  Monnier2005,Kraus2008lines,Kraus2008,Weigelt2011}. {{An extensive
overview of the structure of the inner discs of Herbig Ae/Be stars is
presented in \citet{DullemondandMonnier2010}. Here, we focus on the differences between the
circumstellar environments of Herbig Ae (HAe) and Herbig Be (HBe)
stars.}}

\smallskip

Based on the analysis
of the infrared excesses of such objects, \citet{Meeus2001} suggest
that the disks of HAe/Be stars can be split into two Groups: I \&
II. It has been proposed that these two groups represent different
disc geometries. Group I objects, objects with prominent mid infrared
excesses, are thought to possess flared discs. Group II objects, which
have less strong excesses in the mid infrared, are thought to possess
flatter disc geometries. While the majority of HAe stars are
classified as Group I objects, HBe stars generally belong to Group II
\citep{Acke2005}. Whether this dichotomy is due to the more rapid
evolution of luminous YSOs, a consequence of the dependence of disc
geometry on the temperature of the central star or a combination of
these and other factors is not clear.

\smallskip

Another difference between the discs around HAe and HBe objects was
discovered with long baseline interferometry in the infrared. In the
past decade, the circumstellar environments of many HAe/Be objects
have been spatially resolved by interferometric observations in the
near infrared \citep[see e.g.][]{Millan-Gabet1999,Millan-Gabet2001,
  Eisner2003,Eisner2004,Eisner2007,Eisner2009,Eisner2010,Monnier2005,Monnier2006,Malbet2007,
  Kraus2008,Tannirkulam2008, Benisty2010,Renard2010,Kreplin2012}. It
has been noted that low and intermediate luminosity HAe and HBe
objects follow a tight correlation between their size in the $K-$band
and their luminosity. However, it has also been found that the most
luminous HBe objects appear undersized based on this relationship
\citep{Monnier2005}. It has been proposed that this could be due to
the presence of an optically thick inner disc that shields the outer
dust disc from stellar radiation, allowing it to exist closer to the
central star \citep{Monnier2002}. {{This inner disc may represent an
optically thick accretion disc, and this hypothesis is consistent with
broad-band, long baseline interferometry in the infrared
\citep{Monnier2005,Vinkovic2007}}}. Furthermore, such discs can provide
a significant contribution to the NIR excess of their host star. Since
this emission originates from within the dust sublimation radius, it
will also contribute to the object appearing undersized. This
hypothesis has been shown to be valid in a few cases
\citep[see][]{Kraus2008}. However, since only a few high luminosity
HBe stars have been studied with high resolution, the structure and
evolution of their discs is still not fully understood. {{Indeed,
    it is still not clear whether or not all luminous YSOs are
    undersized \citep[see e.g.][]{DullemondandMonnier2010}.}}

\smallskip

The issue is further complicated by the fact that it can be difficult
to determine the evolutionary status of luminous emission line
objects. Several objects that may be Herbig Be stars could also be
evolved objects \citep[see e.g.][]{Kraus2009}. The uncertain
evolutionary status of such objects makes it difficult to obtain an
overview of the circumstellar environment of luminous YSOs.

\smallskip

To address these issues, we observed the Herbig Be candidate HD 85567
using the VLTI and AMBER. HD 85567 (CPD $-$60$^{^{\circ}}$1510, Hen
3-331) is a luminous B[e] object of uncertain evolutionary status. It
is listed as Herbig Be star with spectral type B5Ve by
\citet{The1994}. The object was also listed as a HAeB[e] star, i.e. a
Herbig Be star that exhibits forbidden line emission, by
\citet{Lamers1998}. Based on the object's infrared SED,
\citet{Malfait1998} classify HD 85567 as an object with double dust
disk. However, the lack of a significant dip between the near and mid
infrared flux indicates that the disc of HD 85567 is relatively
un-evolved. Recently, \citet{Verhoeff2012} included HD 85567 in a
sample of HAeBe stars observed in the $N-$band. These authors classify
the SED of HD 85567 as that of a type II object using the scheme of
\citet{Meeus2001}. This could indicate the presence of an optically
thick inner disc that prevents the outer disc from flaring.

\smallskip

The numerous studies mentioned above classify HD 85567 as a relatively
luminous ($L_{\star}\sim15\,000~L_{\odot}$) YSO. However, there is an
alternative scenario. \citet{Miro2001} note that HD 85567 does not
exhibit a prominent far infrared excess. These authors suggest that
this could indicate that this object is not a YSO and that the
presence of warm dust in its environment might be attributed to mass
loss driven by binary interactions. However, this was only a
conjecture as these authors do not find direct evidence of a
companion. HD 85567 was later classified as a binary by
\citet{DB2006}. These authors present a clear photo centre shift of 29
milli-arcsec to the south over the H$\alpha$ line. Assuming that the
flux ratio in the optical is unity and that only one component
exhibits H$\alpha$ emission, this implies a separation of around 60
milli-arcsec (mas), approximately 100~AU at 1.5~kpc. However, it is
likely the optical flux ratio is not unity and thus that the
separation is many times larger. Therefore, it is not clear if the
companion detected is close enough to interact with the primary, thus
causing the B[e] behaviour of this object. As a result, at present, it
is still not clear whether HD 85567 is a bone fide YSO or an
interacting B[e] binary.

\smallskip

High resolution observations have the potential to address this issue
by examining whether HD 85567 has an additional companion at a smaller
separation. Furthermore, high resolution observations also provide an
opportunity to probe the structure of the object's disc. With this in
mind, we observed HD 85567 with the VLTI and AMBER. This paper
presents the observations and is structured as follows. The
observations are presented in Sect. \ref{SECT:OBS_AND_DATA}. The
results are presented in Sect. \ref{SECT:RESULTS} and discussed in
Sect. \ref{SECT:DISC}. The paper is concluded in
Sect. \ref{SECT:CONC}.

\section{Observations and data reduction}
\label{SECT:OBS_AND_DATA}

\begin{center}
  \begin{figure}
    \begin{center}
      \includegraphics[width=0.4\textwidth]{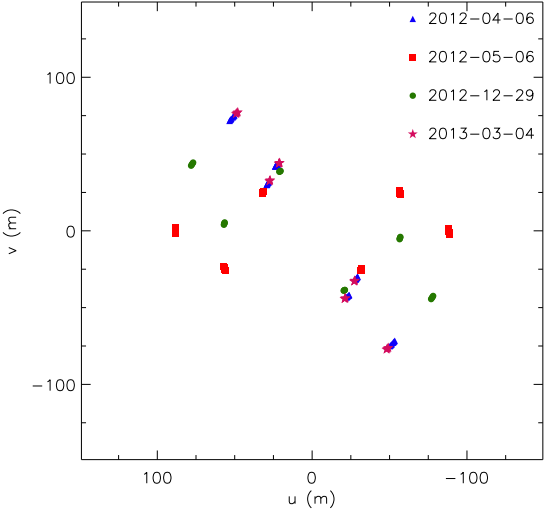}
      \caption{Projected baselines of the observations of HD 85567.\label{FIG:UV}}
    \end{center}
  \end{figure}
\end{center}

HD 85567 was observed with the VLTI and AMBER in the $K-$band using the medium spectral resolution mode. This provides a spectral resolution of $R \sim 1500$ or $\Delta v \sim$200$\mathrm{km\,s^{-1}}$ and a wavelength coverage of $\sim$2.15--2.45~$\mu$m. Observations were conducted using the UT1-UT2-UT3 telescopes on two occasions and the UT2-UT3-UT4 telescopes on two additional occasions. In all cases, {{FINITO \citep{FINITO}}} was used to provide fringe tracking. The observations span a period of approximately 11~months. In all cases, observations of HD 85567 ($H = 6.7$, $K = 5.8$) were conducted between observations of the calibrator objects HD 85313 ($H = 5.3$, $K = 5.1$) and HD 84177 ($H = 5.4$, $K = 5.3$). The projected baselines are displayed in Fig. \ref{FIG:UV} and a log of the observations is presented in Table \ref{TAB:LOG}.

\smallskip

The data were reduced in the standard fashion for AMBER data using the JMMC amdlib package\footnote{Version 3.0.3, available at http://www.jmmc.fr/amberdrs} \citep[see][]{Tat-amdlib,Chelli2009}. A variety of {{selection rates were used to choose frames of the interferograms for processing. Accurate visibilities require a low selection rate, as low S/Ns can bias the results, while precise differential phases can benefit from relatively high selection rates. They are not biased in the same and thus the precision can be increased by increasing the amount of frames selected.}} Calibration of the data, visibilities and closure phases, was performed using a transfer function constructed from the observations of the calibrators. The transfer functions were constructed assuming that the two calibrators, HD 84177 and HD 85313, can be described as uniform discs with radii given by $0.435 \pm 0.031$ \& $0.449 \pm 0.032$~mas respectively \citep{VLTI_Cal_1,VLTI_Cal_2}. 

\begin{center}
  \begin{table*}
    \begin{center}
      \caption{Details of the series of AMBER observations.\label{TAB:LOG}}
      \begin{tabular}{l l l l l l l l }
        \hline
        \hline
        Object & Seeing & Coherence time &  DIT$^1$ & Telescopes & Baselines & PAs \\
              &(\arcsec) & (ms) & (ms) & & (m) & (${\degr}$) \\
        \hline

 \multicolumn{5}{l}{\bf{{2012/04/06}}}\\
        HD 84177 & 0.81 & 2.98 & 500.0 &  UT1-UT2-UT3 & 50.0/43.3/92.6 & 23.8/37.7/30.3\\
        HD 85567 & 0.78 & 3.10 & 500.0 &  UT1-UT2-UT3 & 48.4/42.1/89.8 & 28.5/43.1/35.2\\
        HD 85313 & 0.78 & 3.11 & 500.0 &  UT1-UT2-UT3 & 48.0/41.5/88.7 & 32.1/47.3/39.1\\

        \multicolumn{5}{l}{\bf{2012/05/06}}\\
        HD 84177 & 0.88 & 5.60 & 500.0 & UT2-UT3-UT4& 41.8/89.0/61.3 & 226.7/84.2/108.8 \\
        HD 85567 & 1.06 & 4.95 & 500.0 & UT2-UT3-UT4& 40.5/88.3/61.8 & 231.7/89.8/113.6\\
        HD 85313 & 0.55 & 8.52 & 500.0 & UT2-UT3-UT4& 37.6/85.5/62.5 & 242.9/104.4/127.9\\

        \multicolumn{5}{l}{\bf{2012/12/29}}\\
        HD 84177 & 0.55 & 4.11 & 500.0 & UT2-UT3-UT4& 44.5/88.4/56.1 & 205.7/58.2/83.4\\
        HD 85567 & 0.52 & 4.35 & 500.0 & UT2-UT3-UT4& 44.0/88.9/56.9 & 208.2/60.7/85.3\\
        HD 85313 & 0.70 & 3.31 & 500.0 & UT2-UT3-UT4& 43.8/89.2/58.0 & 211.9/65.6/90.3\\
        \multicolumn{5}{l}{\bf{2013/03/04}}\\
        HD 84177 & 0.86 & 4.33 &  500. 0 &  UT1-UT2-UT3&50.0/43.3/92.6 & 23.7/37.6/30.1\\
        HD 85567 & 0.55 & 5.80 &  500.0 & UT1-UT2-UT3 & 48.9/42.6/90.8 & 25.8/40.0/32.4\\
        HD 85313 & 0.54 & 6.64 &  500. 0 &  UT1-UT2-UT3& 48.7/42.2/90.1 & 28.9/43.6/35.7\\
        \hline
        \end{tabular}
        \tablefoot{$^1${{DIT represents the Detector Integration Time.}}}
    \end{center}
  \end{table*}
\end{center}

Comparison of the transfer functions and observations of HD 85567
(shown in Fig. \ref{FIG:TRANS}) reveals an apparent change in the
appearance of the target. In two cases (2012/04/06 and 2013/03/04),
the visibilities of HD 85567 are the same as the calibrators,
indicating a compact source. In the other two cases (2012/05/06 and
2012/12/29), the visibilities of HD 85567 are noticeably lower than
the calibrators, indicating an extended source. If this behaviour were
real, this would indicate that the environment of HD 85567 was compact
at the beginning of our observations, became extended and then
returned to its initial appearance.

\smallskip

 The simplest explanation of this behaviour is that HD 85567 has a
 previously undetected binary companion and the period of the system
 is of the order of approximately 1 year. However, the lack of a
 strong closure phase signal is not consistent with this
 scenario. Therefore, we explored the possibility that an
 observational bias is affecting the visibilities (this is discussed
 in App. \ref{SECT:APP_B}).

\smallskip

It was found that when the target visibilities are significantly lower
than those of the calibrators, there is a marked difference in the
distributions of ratio of the target and calibrator fringe signal to noise
(S/N). {{The S/N associated with the fringes is the S/N of the
coherent flux, and is an important quantity to consider in the
reduction of AMBER data \citep{Tatulli2007}}}. We surmise that the
difference between the target and calibrator observations on these
dates is a bias caused by the fringe tracking performance degrading
when observing the target. This is supported by the FINITO data
recorded by the RMNREC software. The degradation of FINITO's
performance when observing the target was likely due to two
reasons. Firstly, the target is a magnitude fainter than the
calibrators in the $H-$band where fringe tracking is
conducted. Secondly, the science observations were associated with
poor seeing (especially on 2012-05-06). It is surmised that on the
dates in question, poor fringe tracking resulted in an artificially
lower fringe contrast for the observations of HD 85567, when compared
to the calibrator observations. Consequently, only the observations
when the fringe S/N distributions of the target and calibrator
observations are similar can be calibrated. In principle, the
observations of 2012-04-06 and 2013-03-04 offer reliable
calibration. However, since the fringe S/Ns of the observations
conducted 2012-04-06 are relatively low, the rest of the paper focuses
exclusively on the observations obtained on the date 2013-03-04. These
data were taken after AMBER's performance was improved in January 2013
and thus both the target and calibrator observations exhibit
relatively high fringe S/Ns (see Fig. \ref{FIG:FRAME_SNR}).

\section{Results}
\label{SECT:RESULTS}
\begin{center}
  \begin{figure*}
    \begin{center}
      \begin{tabular}{l l}
         \includegraphics[width=0.43\textwidth]{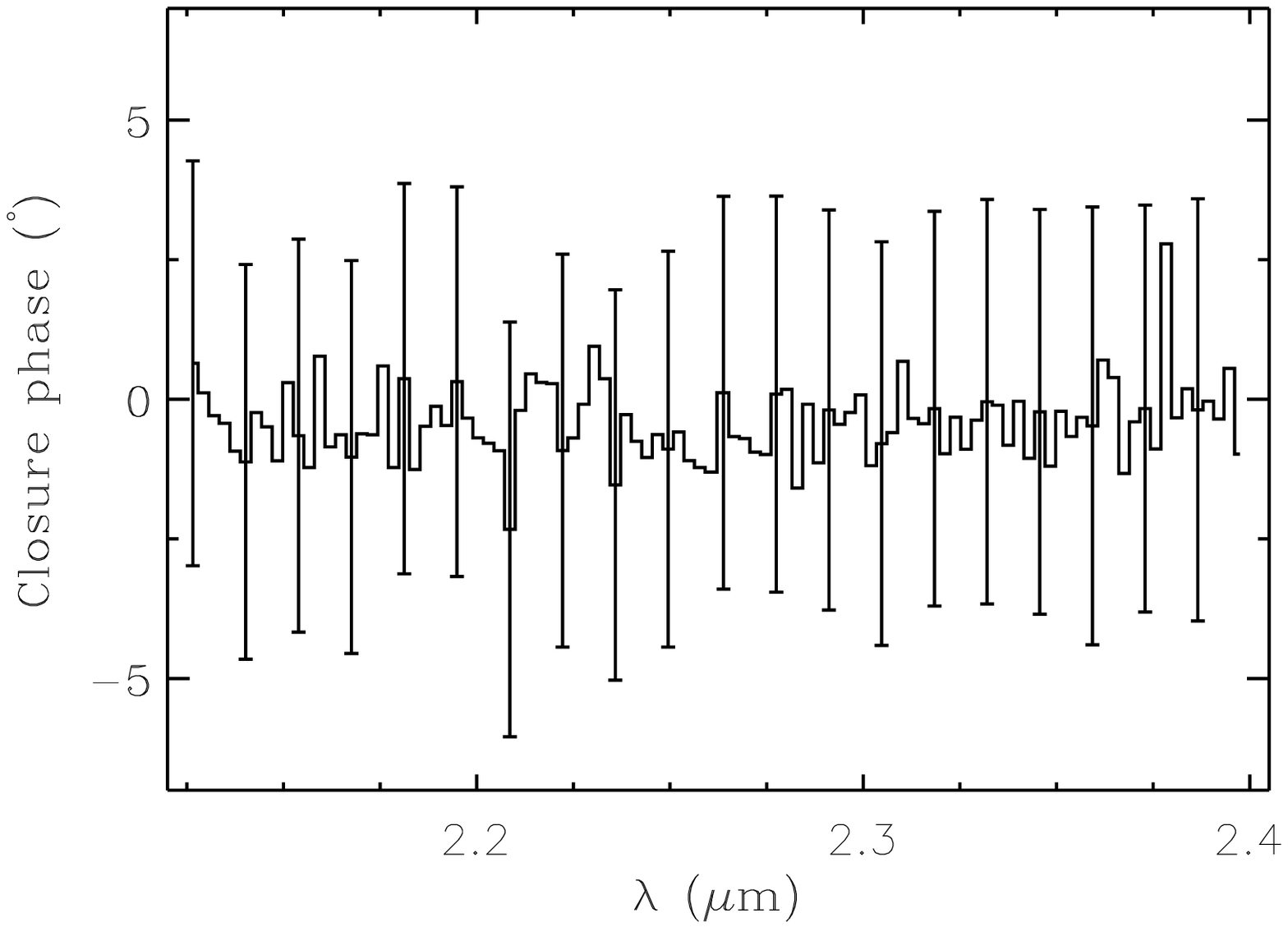} &
        \includegraphics[width=0.45\textwidth]{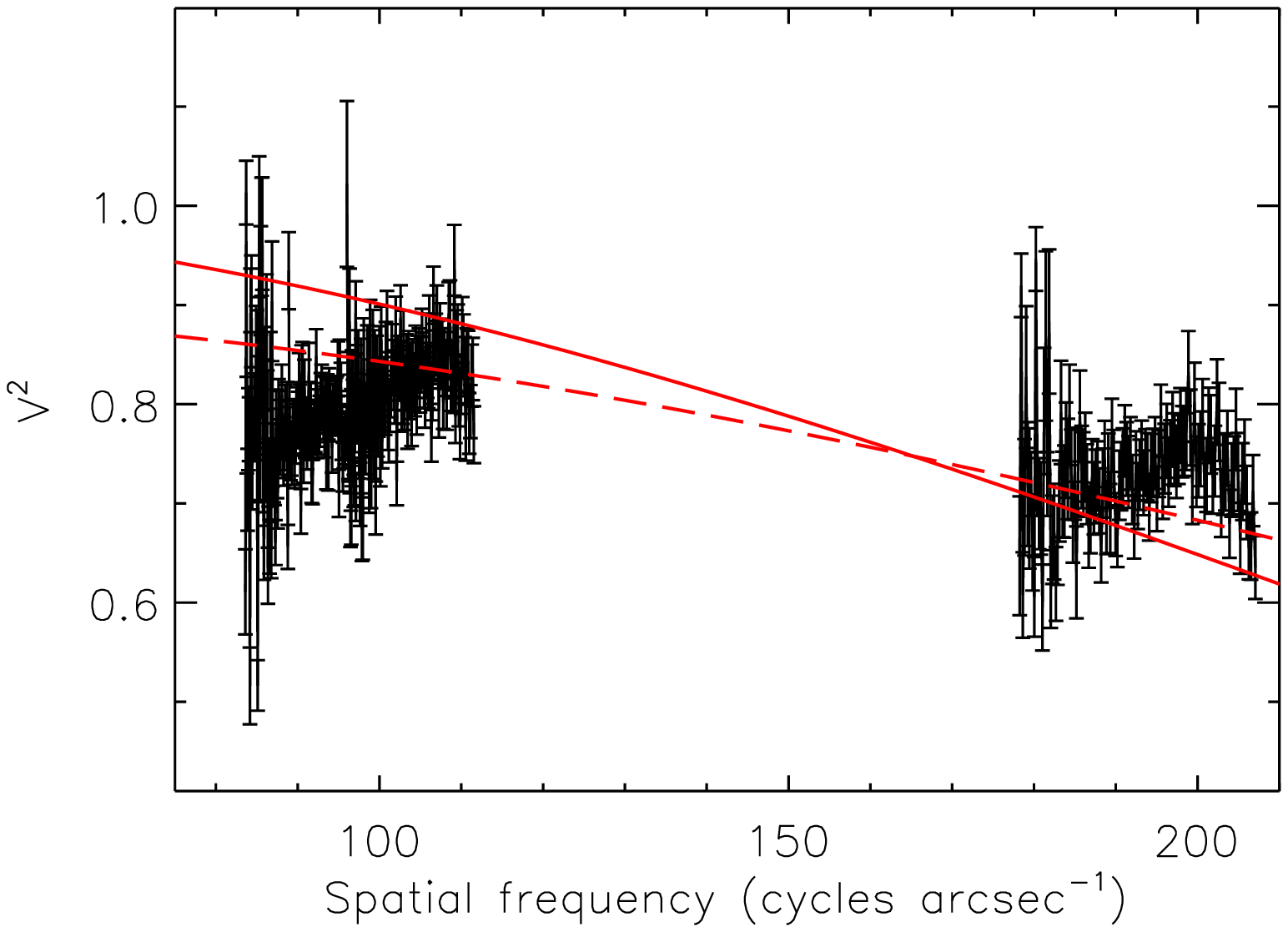} \\

      \end{tabular}
      \caption{Time-averaged closure phase and squared visibility
        observations of HD 85567. The panel on the left presents the
        closure phases. The closure phase error bars shown represent
        the mean error in the measurements. A frame
        selection of 80 per cent was used. In the panel on the right,
        we present the squared visibilities. A frame selection of 20
        per cent was used. The errors represent the mean error in the
        calibrated visibilities. The solid line is the visibility
        profile of a ring with a radius of 0.69~mas, 1.0~AU at
        1.5~kpc. The long-dashed line corresponds to a ring with a
        radius of 0.56~mas (0.8~AU) with the addition of a background
        that accounts for 5 percent of the total
        flux.\label{FIG_V2_CP}}
    \end{center}
  \end{figure*}
\end{center}

The interferometric observations of HD 85567 conducted on 2013-03-04
are presented in Fig. \ref{FIG_V2_CP}. The time averaged closure phase
is close to zero. We conclude that there is no compelling evidence
that the environment of HD 85567 is asymmetric on the scales probed by
these observations. The calibrated visibilities are relatively high,
$\sim0.7-0.8$. This indicates that the environment of HD 85567 is only
marginally resolved. To determine the characteristic size of the
continuum emission region, the calibrated visibilities were fit with a
geometric ring model. This is discussed in Sect. \ref{SECT:VIS_FITS}.

\smallskip

\begin{center}
  \begin{figure*}
    \begin{center}
      \begin{tabular}{l l}
  
        \includegraphics[width=0.35\textwidth]{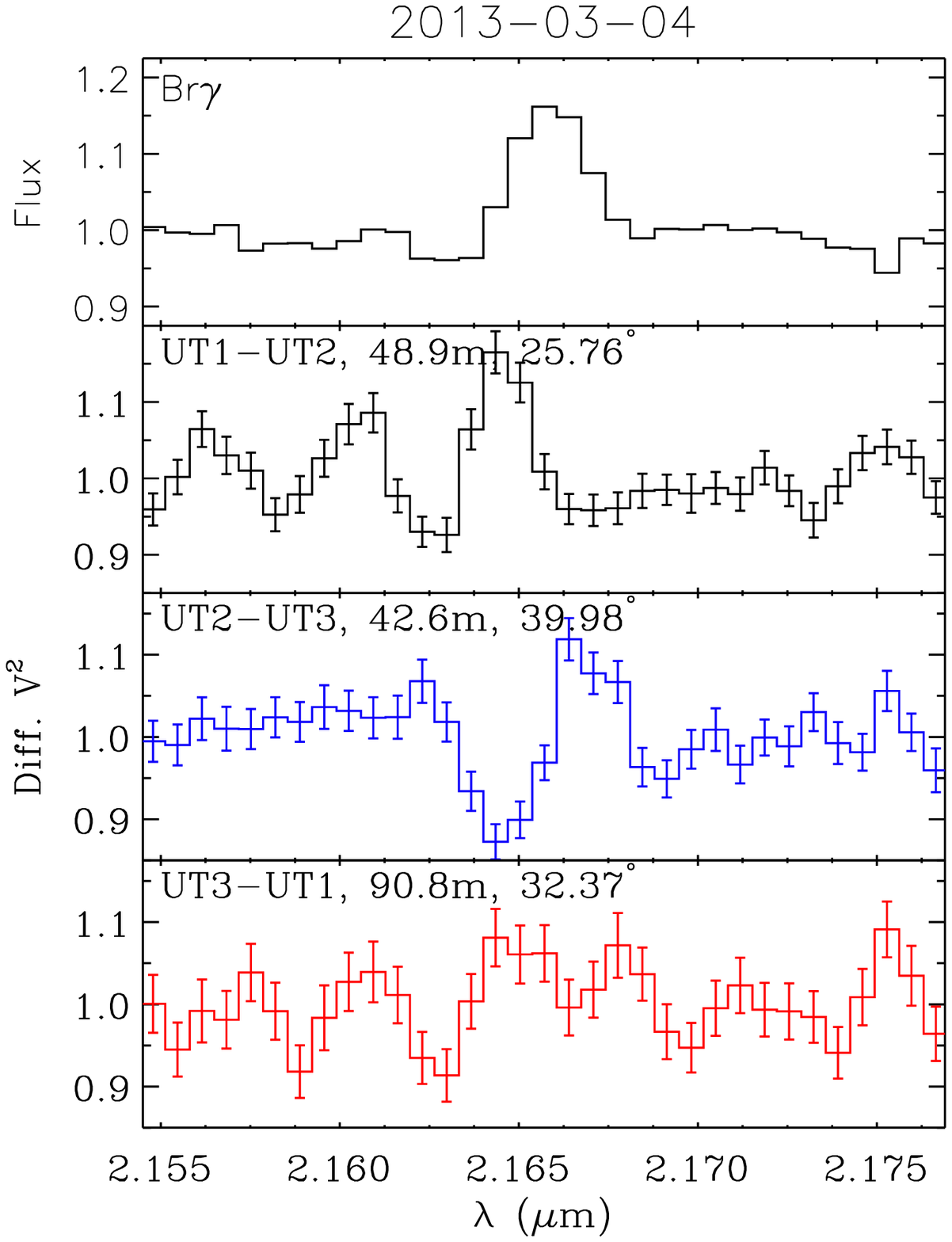} &  

\includegraphics[width=0.35\textwidth]{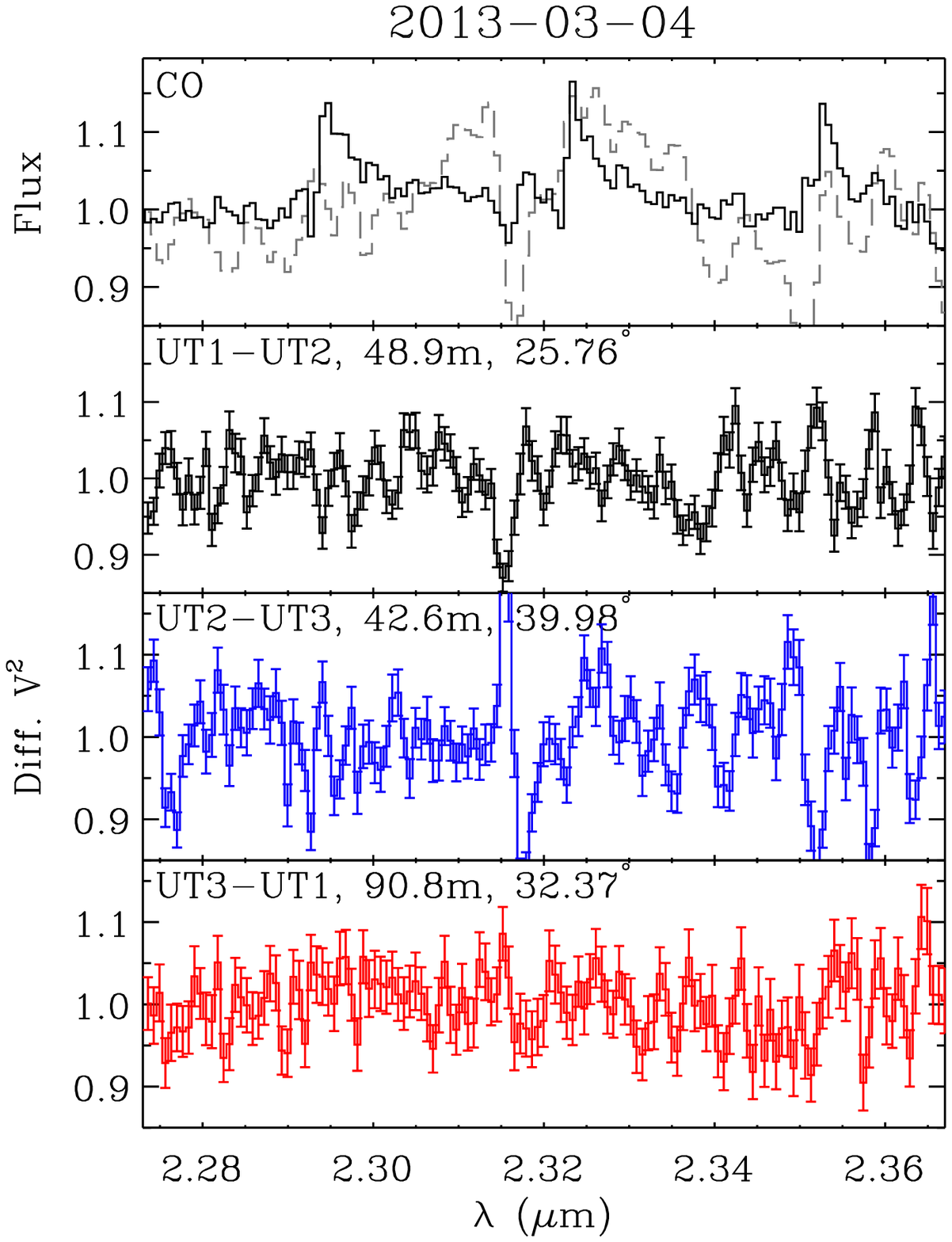} \\

 \includegraphics[width=0.35\textwidth]{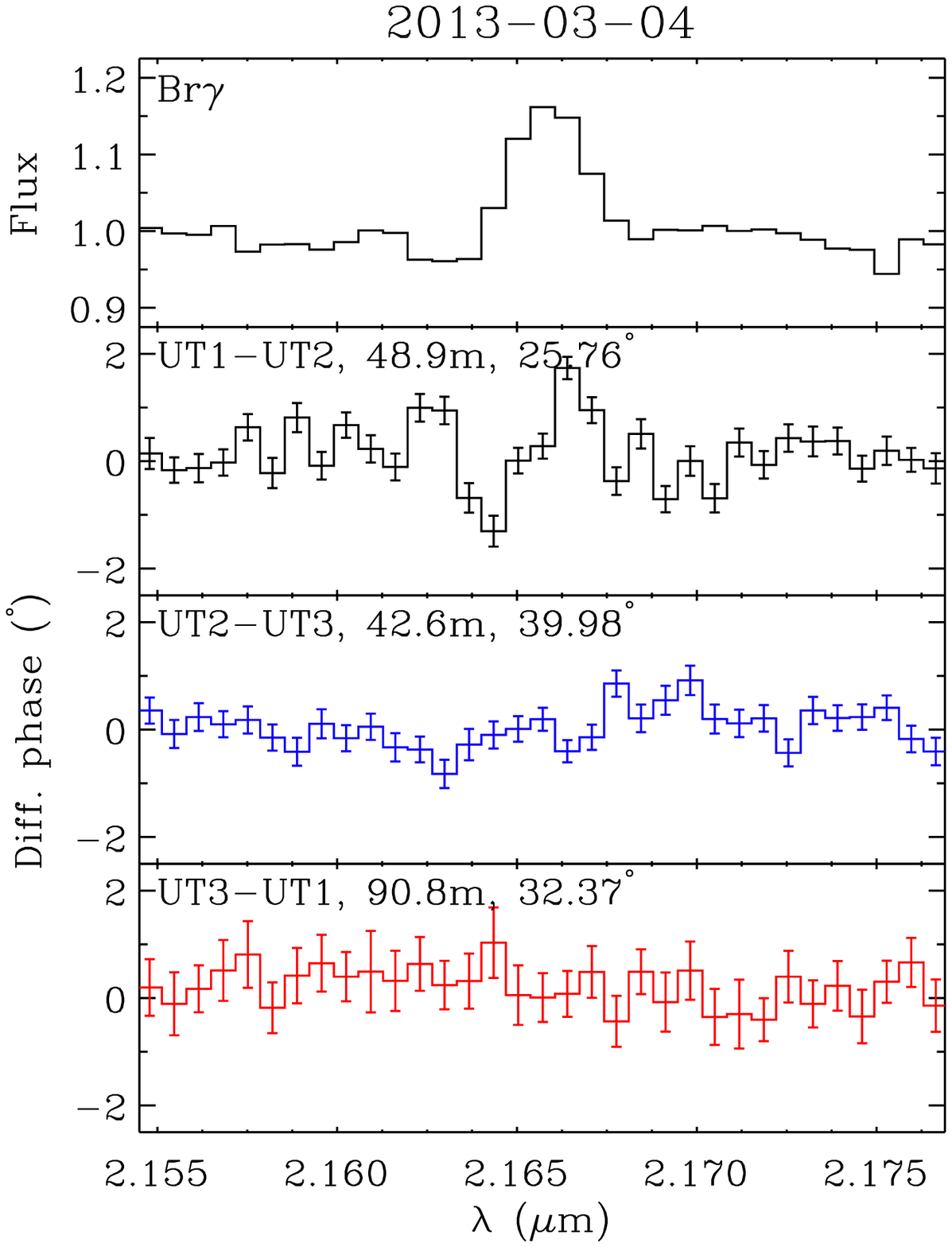} &   
 \includegraphics[width=0.35\textwidth]{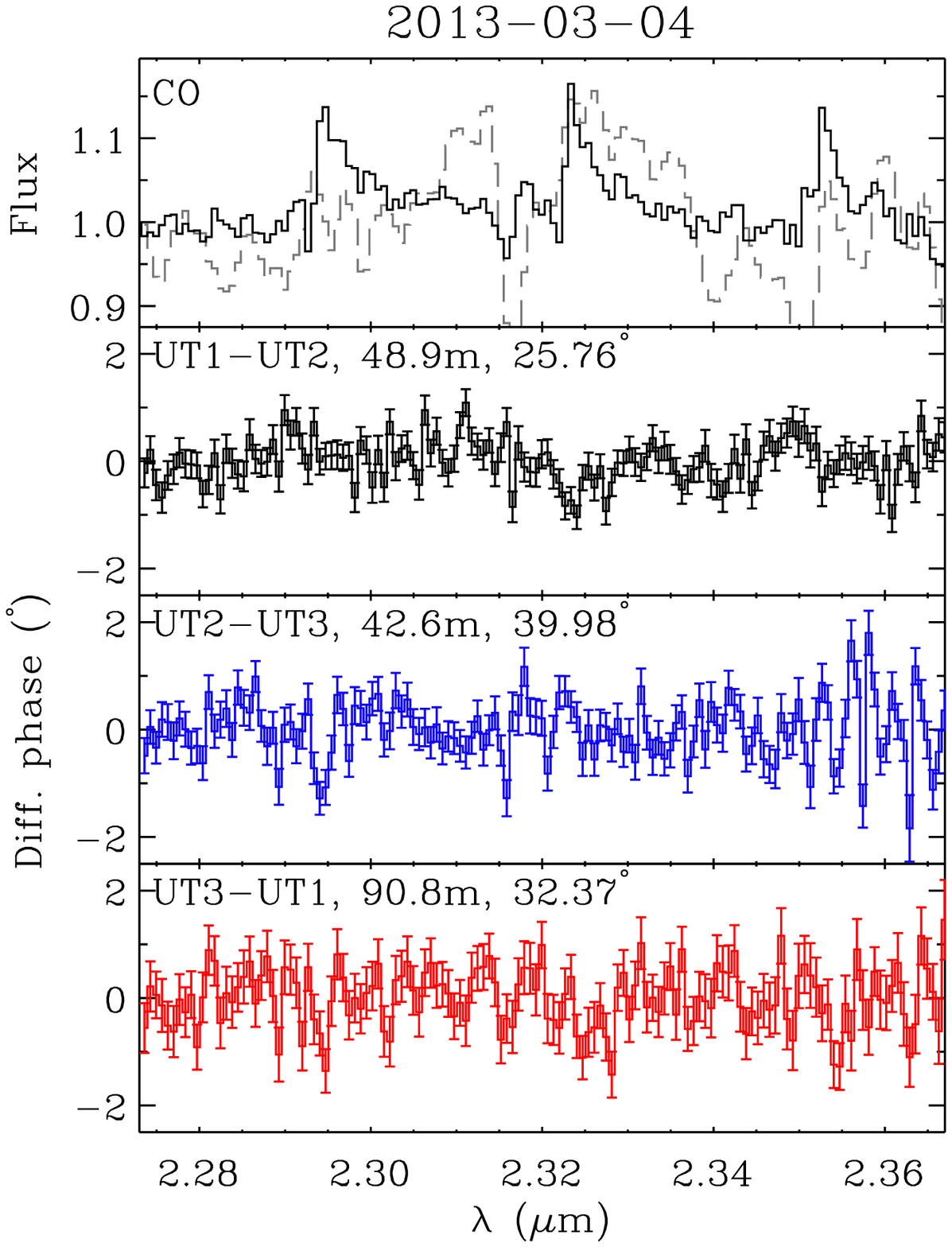} \\

      \end{tabular}
      \caption{Time-averaged differential squared visibility (top) and differential phases (bottom) around the Br$\gamma$ line (left) and CO first overtone bandhead emission (right). Error bars shown represent statistical errors on the mean. A frame selection of 20 per cent was used for the visibilities and a selection rate of 80 percent was employed to obtain the averaged phases. The {{individual AMBER files}} were merged before frame selection was conducted. The dashed line alongside the CO emission represents the spectrum before telluric correction. \label{FIG:DIFF_V2_AND_PHASE}}
    \end{center}
  \end{figure*}
\end{center}

The $K-$band spectrum of HD 85567 exhibits Br$\gamma$ and CO first
overtone bandhead emission. The differential visibilities and phases
across the Br$\gamma$ and CO overtone emission are presented in
Fig. \ref{FIG:DIFF_V2_AND_PHASE}. In both cases, no conspicuous
signature is observed. This suggests that the distributions of the
continuum, Br$\gamma$ and CO overtone emission are similar. However,
it is possible a slight change occurs in the differential phases
associated with the longest baseline over the CO bandhead
emission. This is discussed in more detail in Sect. \ref{SECT:DIFF}.

\smallskip

\subsection{Ring model}

\label{SECT:VIS_FITS}

\begin{center}
  \begin{figure}
    \begin{center}
      \includegraphics[width=0.45\textwidth]{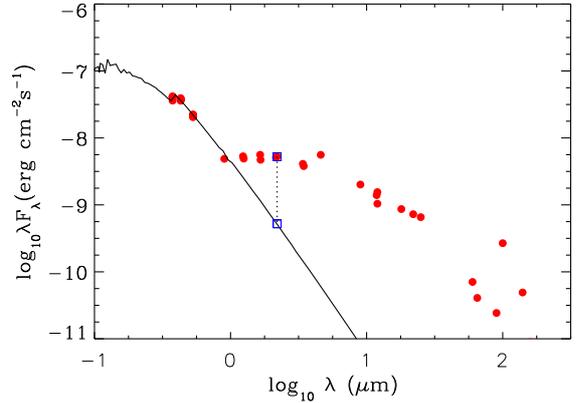}
      \caption{HD 85567's SED. The data were de-reddened using an
        $A_V=1.1$, and are shown with the predicted SED for the parameters $T_{\rm{eff}}$=19\,000 and $\log g$=3.5. \label{FIG:SED}}
    \end{center}
  \end{figure}
\end{center}

To fit the visibilities, the ratio of the infrared excess and
photospheric emission was determined. This was achieved by analysing
the SED of HD 85567, which was constructed by taking data from the
literature and using the VOSA utility \citep{VOSA}. The SED contains
data from 2MASS \citep{2MASS_ref}, AKARI
\citep{Ishihara2010,Murakami2007,Onaka2007}, DENIS \citep{DENIS}, IRAS
\citep{IRASPS,IRAS_2}, Tycho-2 \citep{TYCHO}, WISE
\citep{Wright2010,WISE}, \citet{deWinter2001}, \citet{Klare77},
\citep{Schild83}, \citet{Miro2001}, \citet{Mer1994} and
\citet{Verhoeff2012}. The data were de-reddened using the extinction
relationship of \citet{Cardelli1989} and values of $A_V = 1.1, R_V = 3.1$ (see
Table \ref{TAB:PAR}). The final SED is shown in Fig. \ref{FIG:SED}.

\smallskip

To determine the ratio of the stellar and circumstellar emission, we
compared the observed SED to the stellar flux expected for the
spectral type of HD 85567. The stellar parameters taken from the
literature, including spectral type and effective temperature, are
listed in Table \ref{TAB:PAR}. Assuming an effective temperature of
$T_{\mathrm{eff}} = 19\,000$, the ratio of the excess to stellar
emission in the $K-$band is 9.0.

\begin{center}
  \begin{table}
    \caption{Adopted stellar parameters.\label{TAB:PAR}}
    \begin{center}
      \begin{tabular}{l l l}
        \hline
        \hline
        Parameter & Value & Ref.\\
        \hline
        Spec. Typ. & B2& M01\\
        $T_{\rm{eff}}$ & 19\,000~K& M01\\
        $d$ & $1.5\pm0.5$~kpc& M01\\
        $R_{\star}$ & $9\pm2$~$R_{\odot}$& V12\\
        $A_{\rm{V}}$& $1.1\pm0.1$& V12 \\ 
        $\log L_{\star}$& $4.17\pm0.16$~$L_{\odot}$& V12 \\ 
        $M_{\star}$ &$12 \pm 2$~$M_{\odot}$ & V12\\ 
        \hline
        \end{tabular}
        \tablefoot{M01: \citet{Miro2001}, V12: \citet{Verhoeff2012}}
    \end{center}
    \end{table}
\end{center}

{{The visibilities were then fit with a model of a ring, which was
assumed to be face on as the data do not show PA-related V2 variations
indicative of an asymmetric object.}} The best fit ring radius was found
to be given by $r = 0.69_{-0.23}^{+0.20}$~mas, which resulted in a
minimum chi squared value of $\chi^2_{\mathrm{R}} = 3.59$ (using the
rms of the visibility measurements). It was found that the fit could
be improved by adding a resolved background component. The minimum
contribution from a totally resolved background flux that resulted in
a $\chi^2 < 2$ was determined to be approximately 5 percent of the
total flux. The resulting best fit ring radius was $r =
0.56_{-0.20}^{+0.16}$~mas, which resulted in a minimum chi squared
value of $\chi^2_{\mathrm{R}} = 1.68$. The best fit visibility
distributions are displayed in the right panel of
Fig. \ref{FIG_V2_CP}.

\subsection{Differential visibilities and phases}
\label{SECT:DIFF}

\smallskip

The differential visibilities across the B$\gamma$ line and CO bandhead emission are presented in Fig. \ref{FIG:DIFF_V2_AND_PHASE}. No clear change in visibilities is observed across the Br$\gamma$ line. There are some suggestions of an increase in visibilities over the line, indicating a compact line emitting region. However, these are not considered significant given the lack of consistency of the position of these increases with respect to the line centre. Since baselines with similar lengths and position angles (UT1-UT2 and UT2-UT3) exhibit different changes in visibility in the approximate region of the Br$\gamma$ line, the features observed are considered artifacts. The differential visibilities over the CO bandhead emission exhibit several artificial changes across telluric absorption lines. These make it challenging to detect changes in the visibilities across the CO bandhead emission. 

\smallskip

The differential phases across the Br$\gamma$ and CO bandhead emission
are also presented in Fig. \ref{FIG:DIFF_V2_AND_PHASE}. In the case of
one baseline (UT1-UT2), it appears that there is a change in phase
across the Br$\gamma$ line. The behaviour of the phase variation with
wavelength; a negative change on the blue side of the line and a
positive change over the red side, is similar to that expected in the
case of line emission originating in a rotating medium. However, since
the phases associated with the similar UT2-UT3 baseline (49 and 43~m
at PAs of $26$ and $40^{\circ}$ respectively) do not exhibit this
behaviour, it is suggested that the phase signal discussed is also an
artifact. In general, no prominent offset is observed in the
differential phases across the CO bandhead. However, since the
observed spectrum features several CO overtone transitions, we
could increase the precision of the differential phase
observations by co-adding the data across the individual
transitions. This was done using the data associated with the longest
baseline (UT3-UT1, 91~m) as these observations access the smallest
scales. The results are discussed in the following section
(\ref{SECT:COADD}).

\subsubsection{Photo-centre offset over the CO bandhead emission}

\label{SECT:COADD}

\begin{center}
  \begin{figure}
    \begin{center}
      \includegraphics[width=0.25\textwidth]{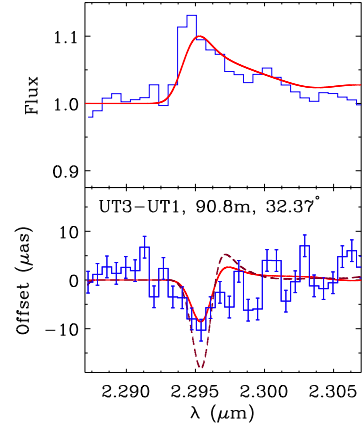}
      \caption{Photo-centre offsets associated with the CO bandhead emission. This figure was constructed by co-adding the time averaged differential phase observations obtained on the date 2013-03-04 using the UT1-UT3 baseline across the first 3 CO bandheads. The uncertainty in the differential phase is represented by the rms of the continuum measurements. The smaller model offset (solid line) corresponds to an inner disc radius of 5 stellar radii (0.2~AU) and an outer radius of 1.0~AU (0.7~mas). The larger offset (dashed line) was calculated for an inner radius of 10 stellar radii (0.4~AU) and an outer radius of 4.0~AU (2.7~mas, sizes calculated assuming a distance of 1.5~kpc).\label{FIG:CO_PHOTO}}
    \end{center}
  \end{figure}
\end{center}

To increase the precision of the differential phase observations
obtained with the UT3-UT1 baseline, the phases across the 3 first
bandhead transitions were averaged. {{The photo-centre offset associated
with the resultant differential phase signal was calculated using $p = -\frac{\phi}{2\pi}\dot\frac{\lambda}{B}$, where $B$ is the baseline length and $p$ represents the projection of the 2D photo-centre along the orientation of the baseline. The result is shown in Fig. \ref{FIG:CO_PHOTO}}}. The observations
are consistent with a small offset corresponding to approximately
10~$\mu$as occurring over the bandhead profile. In contrast, offsets
larger than approximately 10~$\mu$as can be excluded. Whether this can
be used to constrain the location of the CO bandhead emission was then
explored using the model developed in \citet{MeCO,me_hd_327083_2}.

\smallskip

To reduce the running time of the model, it is assumed that the
average photo-centre shift associated with the first three CO overtone
transitions ($2\rightarrow0$, $3\rightarrow1$ and $4\rightarrow2$)
could be modelled as the shift over the first bandhead
($2\rightarrow0$). This is a simplification but ultimately, the
emission of the different bandheads will originate from the same
location. Based on the excitations requirements of the different
transitions, this approach will slightly over-estimate the average
offset. However, given the 0.5~kpc uncertainty in the distance to HD
85567, this was not considered significant.

\smallskip

To calculate the photo-centre offsets associated with CO bandhead
emission from a circumstellar disc, we used the model presented in
\citet{me_hd_327083_2} and the stellar parameters presented in Table
\ref{TAB:PAR}. The source of the CO emission was represented by a
Keplerian disc with power laws describing the radial dependence of the
excitation temperature and surface number density. The exponents of
the respective power laws were set to $p = -0.5$ and $q =
-1.5$. Finally, the inclination was set to $i = 35^{^{\circ}}$, which
is based on fits to the CO bandhead emission presented
in \citep{Ilee_phd}. It is noted that this value is relatively uncertain as it was derived from a model fit to spectra of moderate, rather than high, spectral resolution. Nonetheless, it serves as a representative value and is sufficient for our purposes. {{Once the images of the disc at various wavelengths had been calculated, the associated offset was determined from the photo-centre of each image.}}

\smallskip

We calculated the photo-centre offsets for two
  models. The first with a relatively small inner radius,
  5~$R_{\star}$, and a compact outer radius of 1~AU, as predicted by
  the scenario of an optically thick gas disc interior to the dust
  sublimation radius. The second model featured a larger inner radius,
  10~$R_{\star}$, and a more extended outer radius of 4~AU. This outer
  radius corresponds to the scenario of an optically thin inner disc
  and a dust sublimation radius that reproduces the size luminosity
  relationship of intermediate and low luminosity objects.

\smallskip

The model photo-centre offsets are displayed in
Fig. \ref{FIG:CO_PHOTO}. Clearly, the significance of the slight
offset observed is low, the tentatively identified signature is
approximately 4 times the continuum rms. However, it is evident that
the data are consistent with the offset associated with the smaller
disc. Furthermore, the data favour the smaller disc over the larger
disc as the more extended disc results in an offset that is larger
than that observed.

\section{Discussion}

\label{SECT:DISC}

This paper presents new VLTI/AMBER observations of the B[e] star HD
85567. Two scenarios have been proposed to explain the B[e] behaviour
of this object. One scenario that explains the object's infrared
excess and line emission is that it is a YSO with a circumstellar
accretion disc. The alternative scenario is that HD 85567 is an
interacting binary with circumstellar material that has been deposited
though mass loss driven by binary interactions. Here, we discuss our
findings in the context of these two scenarios. We also briefly discuss the
structure of HD 85567's circumstellar material and consider how this
is evolving.

\smallskip

We note that our moderate spectral resolution observations reveal that
HD 85567 exhibits $^{12}$CO bandhead emission, but not $^{13}$CO
bandhead emission. In principle, the fact that the circumstellar
material of HD 85567 is not significantly enriched in $^{13}$CO
favours the YSO scenario \citep{Kraus2009}. However, we note that
while the spectrum excludes ratios of $^{12}$CO/$^{13}$CO below approximately 15, this is not sufficient to place strong constraints
on the evolutionary status of HD 85567 \citep{Kraus2009}.

\smallskip

The closure phase observations provide an additional means to
investigate the interacting binary hypothesis. HD 85567 has already
been shown to be a binary, although the estimated minimum separation
is $\sim$100~AU \citep[and likely many times this,][]{DB2006}. Since
this companion may be too distant to induce mass loss from the
primary, we used our high resolution observations to investigate the
hypothesis that HD 85567 has an additional, closer companion within
the field of view of the UT telescopes (60~mas, $\sim$100~AU). Since
no closure phase signature is detected, the observations do not reveal
an additional close binary companion. For completeness, we note that
the $u,v$ coverage of the observations discussed is relatively
linear. In principle, a companion could escape detection if it was
aligned perpendicularly to the projected baselines. {{However, the
    additional closure phases associated with the nights of degraded
    FINITO performance are also consistent with zero, and thus
    indicate a symmetric source. This is a robust result as a bias in
    visibilities will not affect closure phase
    measurements}}. Therefore, the data support the conclusion that HD
85567 does not appear to have a close binary companion, although a
faint companion could still escape detection. We now investigate
whether the data are consistent with the hypothesis that HD 85567 is a
YSO.

\smallskip

The observed visibilities are relatively high and can be reasonably
reproduced using a point source and a ring model. We report that the
apparent radius of the $K-$band continuum emitting region of HD 85567
is $r = 0.56_{-0.20}^{+0.16}$~mas ($\sim$$0.8\pm0.3$~AU). Based on the
luminosity of HD 85567 and the predicted dust sublimation radius when
the inner disc is optically thin, the expected ring radius is
4.2~AU. Therefore, this is considerably smaller than expected based on
the size luminosity relationship exhibited by YSOs of low and
intermediate luminosity \citep{Monnier2005}. This is a robust result
as it is most likely independent of a possible bias in the calibrated
visibilities due to the use of FINITO. As discussed previously, FINITO
can bias the target visibilities to low values. Therefore, if the data
are biased, the true size of HD 85567 may be smaller, but not
larger. Furthermore, HD 85567 appears undersized even when allowing
for the uncertainties in its distance and luminosity (both
approximately 30 percent). The undersized appearance of HD 85567 is
similar to the case of luminous YSOs. For example, the Herbig Be star
V1685 Cyg has a luminosity of 21$\,$400~$L_{\odot}$ and a $K-$band
ring fit radius of $2.15^{+0.23}_{-0.18}$~AU, making it undersized by
nearly 3~AU \citep{Monnier2005}. This was also reported to be the case
for the early B type Herbig Be star MWC 297
\citep{Weigelt2011}. Therefore, the size of HD 85567 supports the
hypothesis that this object is also a YSO.

\smallskip

It has been proposed that the reason for the small sizes of luminous
YSO is that their inner discs are optically thick, shielding the inner
rim of the dust disc from stellar radiation. This can allow the dust
sublimation radius to be located closer to the central star than would
otherwise be the case. The optically thick inner gas disc is associated
with active accretion discs interior to the dust sublimation radius
\citep{Eisner2004,Monnier2005}. Here we explore whether this scenario
is applicable to HD 85567. By considering the combined effect of
stellar irradiation and viscous heating, \citet{Millan-Gabet2001}
present the temperature of an accretion disc as a function of
radius. The equations used are the following:

\begin{equation}
T(r) = (T_{\mathrm{rep}}^4 + T_{\mathrm{acc}}^4)^{\frac{1}{4}}
\end{equation}
in combination with
\begin{equation}
T_{\mathrm{rep}} = T_{\star}\left(\frac{1}{3}^{\frac{1}{4}}\right)\left(\frac{R_{\star}}{r}^{\frac{3}{4}}\right)
\end{equation}
and
\begin{equation}
T_{\mathrm{acc}} = \left(\frac{3GM_{\star}\dot{M}_{\mathrm{acc}}}{8\pi\sigma}\right)r^{\frac{-3}{4}}
\end{equation}
where $\sigma$ is the Stefan-Boltzmann constant, $G$ is the gravitational constant and $\dot{M}_{\mathrm{acc}}$ is the accretion rate. These equations can be used to crudely estimate the expected size of accretion discs by determining the radius where the temperature falls to 1500~K, i.e. the approximate dust sublimation radius.

\smallskip

It has been estimated that HD 85567 accretes material at a rate of
rate of approximately $1 \times 10^{-6}~M_{\odot}\,{\mathrm{yr^{-1}}}$
\citep[based on the object's Br$\gamma$ emission,][]{Ilee_phd}. This
should be sufficient to ensure an optically thick inner disc
\citep[see e.g.][]{Weigelt2011}.  Using this accretion rate and the
parameters in Table \ref{TAB:PAR}, we obtain a predicted dust
sublimation radius of 0.9~AU. This is consistent with the best fitting
ring radius of $0.8 \pm 0.3$~AU. Therefore, it is certainly plausible
that the size of HD 85567 in the $K-$band reflects the presence of an
optically thick disc interior to the dust sublimation radius. This is
supported by the finding that a gaseous disc 1~AU in size is
consistent with the differential phase observations over the CO
bandhead emission. Gaseous discs with radii in excess of 4~AU, the
location of the dust sublimation radius in the case of an optically
thin inner disc, do not reproduce the data well.

\smallskip

We conclude that the observations are consistent with the
hypothesis that HD 85567 is a YSO while they do not support the
interacting, evolved binary scenario. We find that HD 85567 appears
undersized according to the size luminosity relationship of YSOs and
demonstrate that this could be due to the presence of an optically
thick gaseous disc interior to the dust sublimation radius. Finally,
we note that the presence of an optically thick inner disc and the
absence of a far infrared excess suggest that HD 85567 is
photo-evaporating its disc from the outside. Supporting the hypothesis
that this is the fate of discs around Herbig Be stars
\citep{Alonso-Albi2009,Verhoeff2012}.

\section{Conclusion}

\label{SECT:CONC}

This paper presents new VLTI/AMBER observations of the enigmatic B[e]
object HD 85567. Here we reiterate the salient results. 

\smallskip

The object's environment appears compact and symmetric on scales of a
few to 100~AU. This does not support the hypothesis that the object is an
evolved, interacting binary. The apparent radius of HD 85567's
environment in the $K-$band is found to be $r =
0.56_{-0.20}^{+0.16}$~mas ($\sim$$0.8\pm0.3$~AU). This makes the
object undersized according to the size luminosity relationship based
on YSOs of low and intermediate luminosity. This has previously been
found to be the case for luminous YSO and thus the size of HD 85567 is
consistent with the hypothesis that it is a YSO.

\smallskip

We then investigate why HD 85567 appears undersized according to the
size luminosity relationship of YSOs. The size of the $K-$band
emitting region is congruous with the predicted location of the dust
sublimation assuming an accretion disc that is optically thick in the
inner regions. Furthermore, the differential phase observations over
the CO bandhead are also consistent with a compact ($r \sim 1$~AU)
gaseous disc interior to the dust disc. More extended discs do not
reproduce the data as well.

\smallskip

To conclude, the data support the hypothesis that HD 85567 appears
undersized according to the YSO size luminosity relationship due to
the presence of an optically thick gaseous disc interior to the dust
sublimation radius. This indicates that HD 85567 is indeed a YSO. If
this is the case, the gaseous inner disc may be identified as an
accretion disc. The presence of an optically think inner disc and the
absence of a far infrared excess suggest that HD 85567 is
photo-evaporating its disc from the outer edge.

\begin{acknowledgements}
  HEW acknowledges the financial support of the Max Planck
  Society. This research has made use of the \texttt{AMBER data
    reduction package} of the Jean-Marie Mariotti Center. This
  publication also makes use of VOSA, developed under the Spanish
  Virtual Observatory project supported from the Spanish MICINN
  through grant AyA2008-02156. 
\end{acknowledgements}

\bibliographystyle{aa} 
\bibliography{bib_2}

\appendix

\section{Transfer functions}

\label{SECT:APP_A}

\begin{center}
  \begin{figure*}
    \begin{center}
      \begin{tabular}{l l }
      \includegraphics[width=0.5\textwidth]{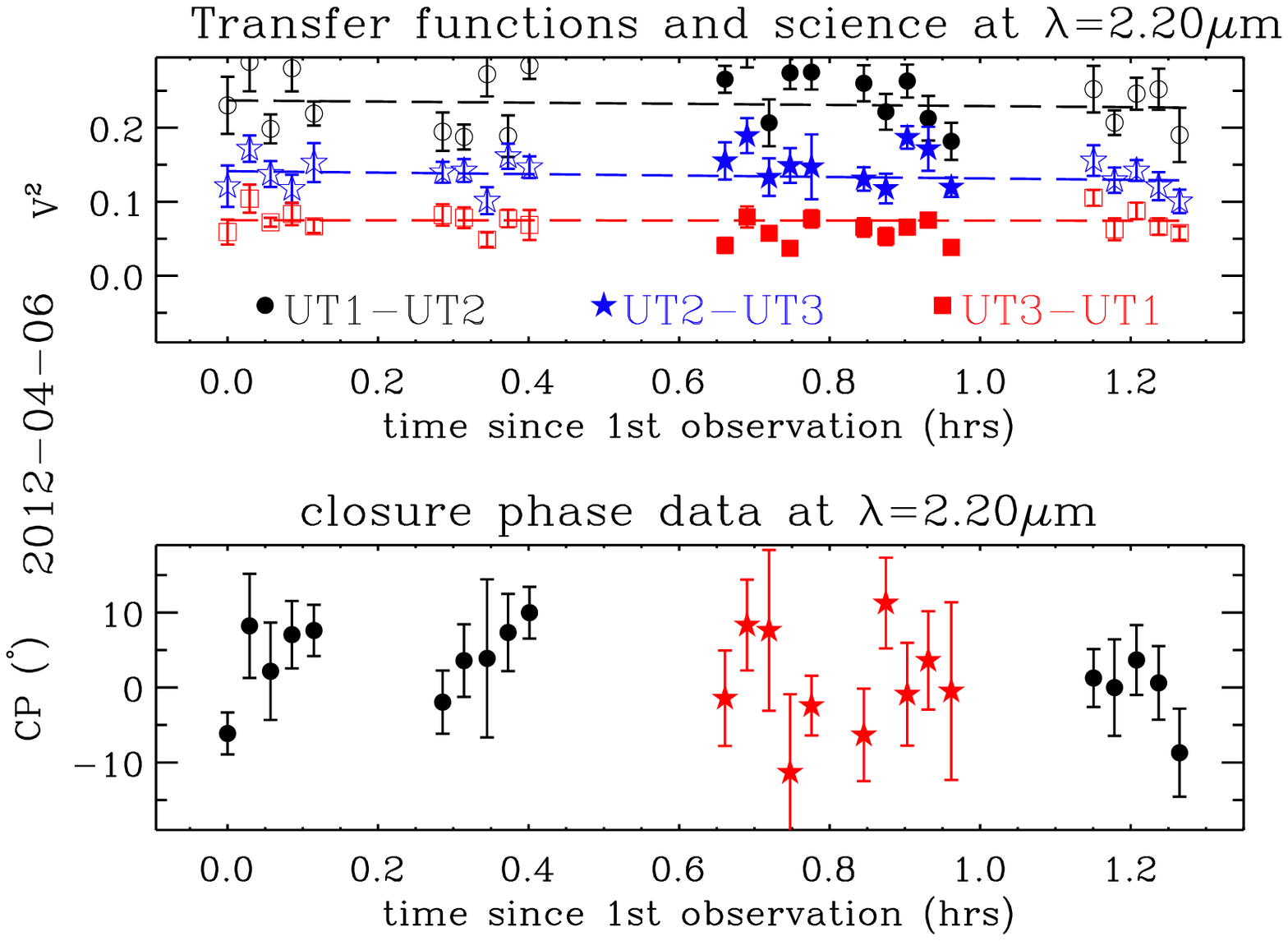} & 
      \includegraphics[width=0.5\textwidth]{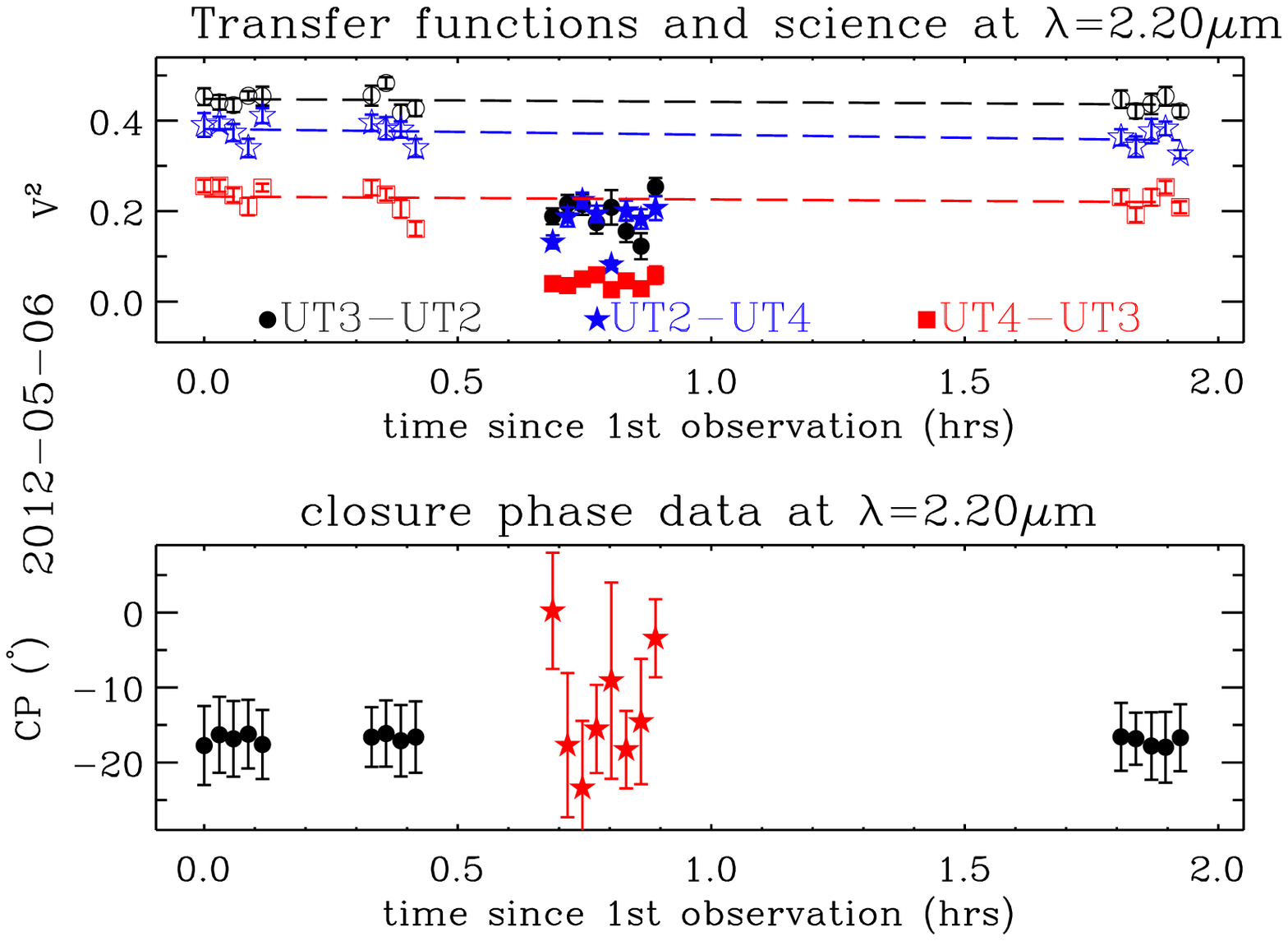} \\ 

      \includegraphics[width=0.5\textwidth]{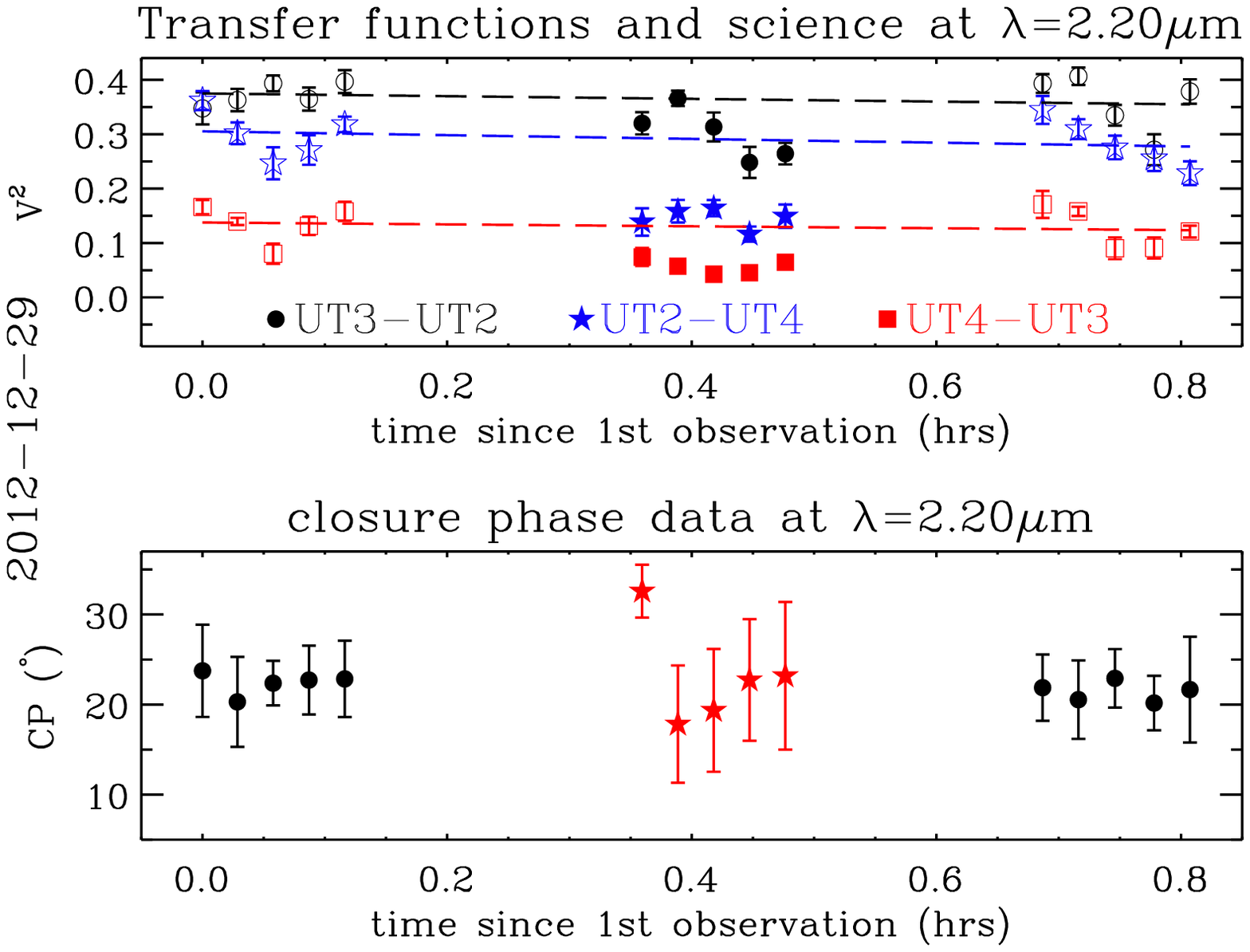} &
      \includegraphics[width=0.5\textwidth]{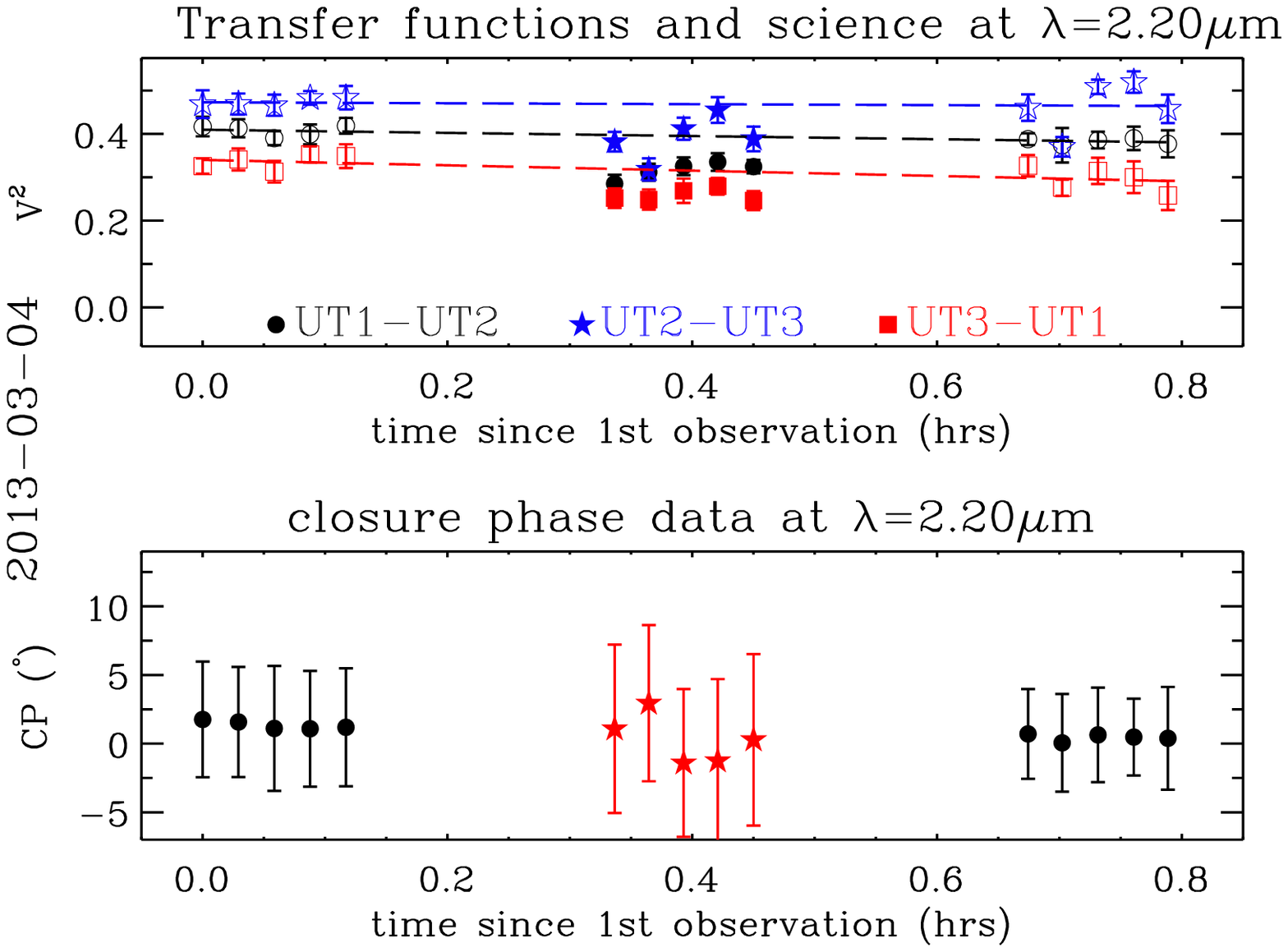}\\
      \end{tabular}
      \caption{Examples of the transfer functions and raw science data at a wavelength of $\lambda=2.2~{\mathrm{\mu m}}$. In the visibility plots, open symbols mark calibrator observations and filled symbols mark the observations of HD 85567. In the closure phase plots, the observations of HD 85567 are marked by star symbols and calibrator data by filled circles. A frame selection of 10 percent was used, but the behaviour is independent of the frame selection used.\label{FIG:TRANS}}
    \end{center}
  \end{figure*}
\end{center}

An example of the transfer functions at a particular wavelength is presented in Fig. \ref{FIG:TRANS}. It can be clearly seen that in two cases (2012/04/06 and 2013/03/04), the visibilities of HD 85567 are the same or similar to the calibrators, indicating a compact source. In the other two cases (2012/05/06 and 2012/12/29), the visibilities of HD 85567 are noticeably lower than the calibrators, indicating a resolved source.

\section{Fringe S/N distributions}
\label{SECT:APP_B}

\begin{center}
  \begin{figure*}
    \begin{center}
      \begin{tabular}{l l }
      \includegraphics[width=0.3\textwidth]{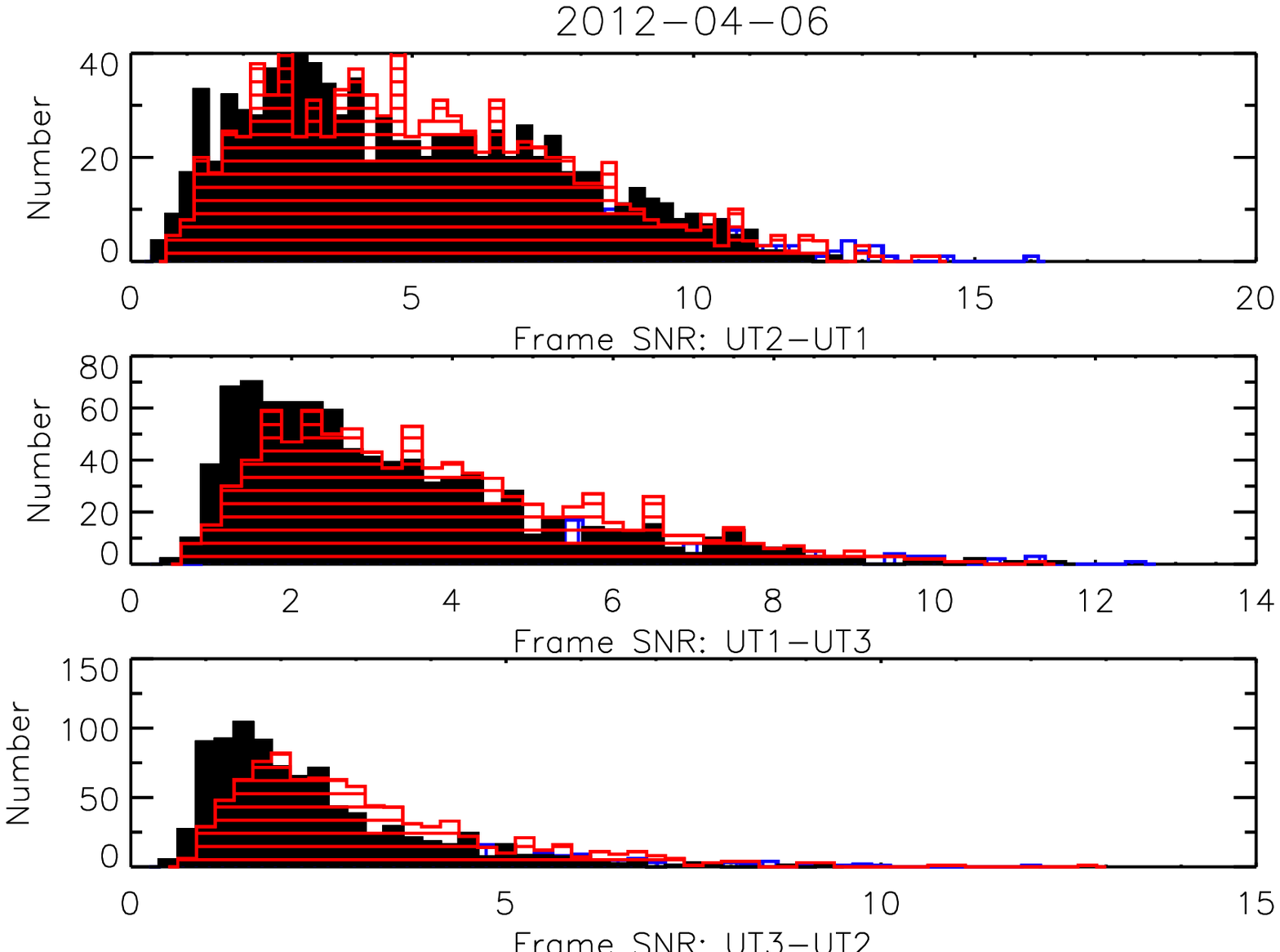} & 
      \includegraphics[width=0.3\textwidth]{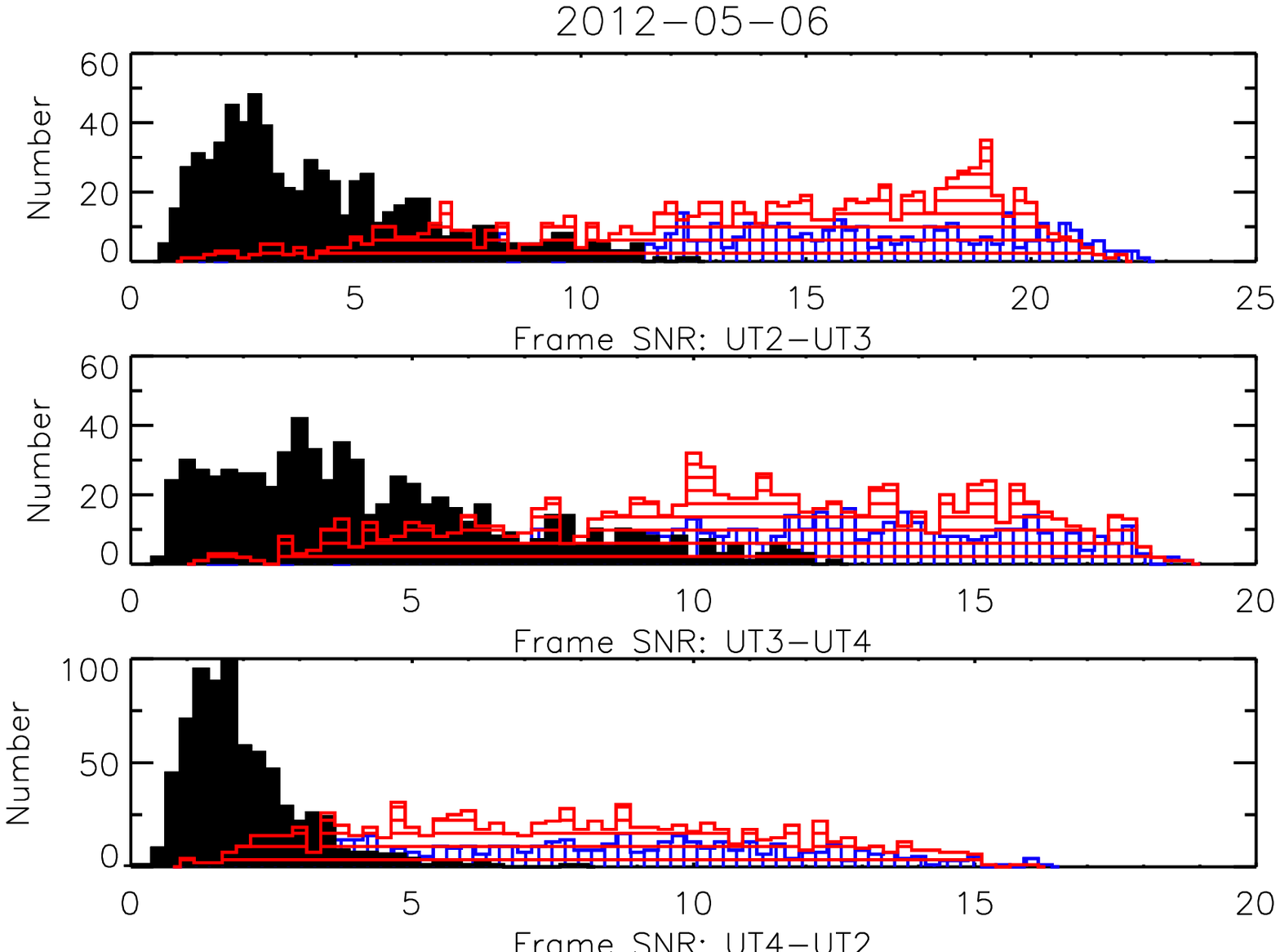} \\

      \includegraphics[width=0.3\textwidth]{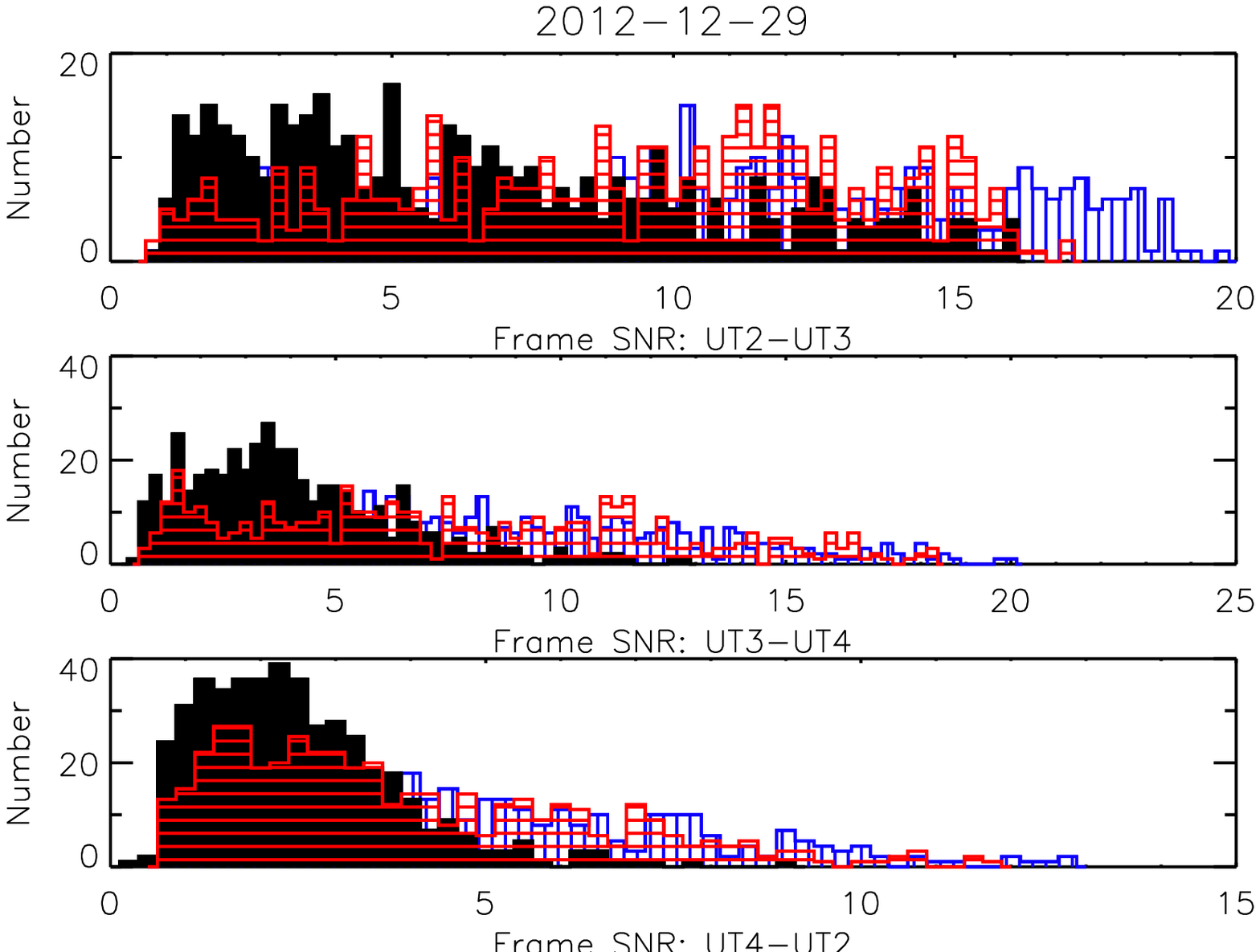} &
      \includegraphics[width=0.3\textwidth]{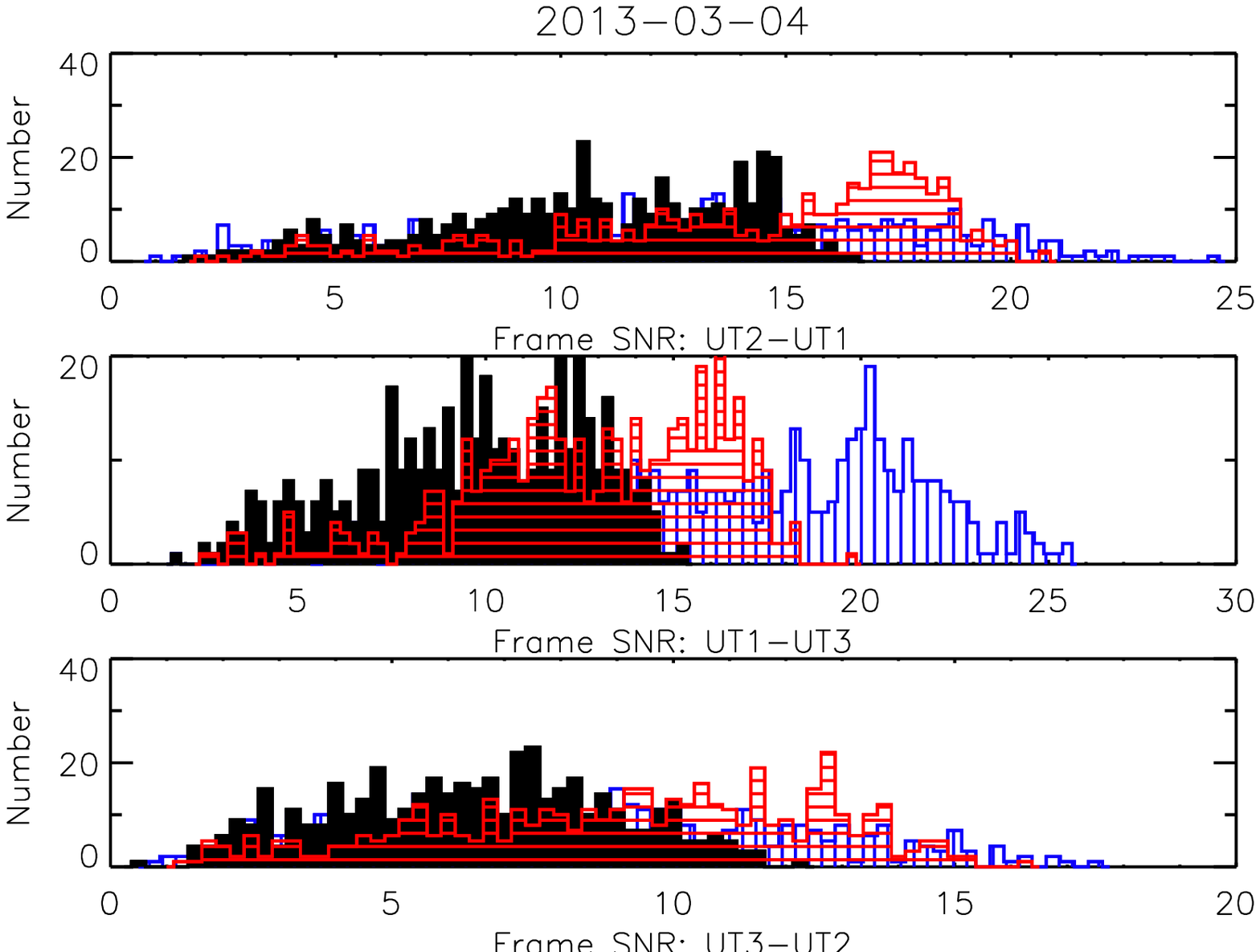}\\
      \end{tabular}
      \caption{Fringe S/N distributions for each observing date and each baseline. The filled black histogram is the fringe S/N distribution of the observations of HD 85567. The distributions filled with vertical blue and horizontal red lines mark the fringe S/Ns associated with the calibrators HD 85313 and HD 84177 respectively.\label{FIG:FRAME_SNR}}
    \end{center}
  \end{figure*}
\end{center}

The distribution of the fringe S/Ns of the science and calibrator
observations are presented in Fig. \ref{FIG:FRAME_SNR}. The two cases
where HD 85567 appears resolved (2012-05-06 \& 2012-12-29) are
considered first.

\smallskip

It can be seen that in these cases, the S/N distributions of the
calibrators extend to significantly higher values than the
distribution of the HD 85567 observations. This is particularly
apparent in the case of the observations conducted on 2012-05-06. On
this occasion, the fringe S/N distribution of HD 85567 on the UT2-UT3
baseline peaks at $\sim$2 while the HD 84177 distribution has a skewed
distribution peaking at $\sim$18. This disparity could be due to the
fact that although the $K-$band magnitudes of the target and
calibrators are similar, HD 85567 is 1 magnitude dimmer than the
calibrators in the $H-$band. The fringe tracking is performed in the
$H-$band. Therefore, the performance of the fringe tracking was likely
higher when observing the calibrators than when observing the
target. Furthermore, the observations of the target were associated
with worse seeing than the calibrator observations. This could also
reduce the fringe tracking performance. Consequently, the high S/N of
the calibrator observations is likely due the fringe tracking
performing best when observing the calibrators. This hypothesis is
substantiated by the FINITO data recorded by the RMNREC software. In
the two cases where the target visibilities appear significantly lower
than those of the calibrators, the rms of the FINITO phases associated
with the target is up to 40 percent larger than that associated with
calibrator observations.

\smallskip

The case of the two dates when the target appears close to unresolved
(2012-04-06 \& 2013-03-04) are now discussed.

\smallskip

In the case of the observations conducted on 2012-04-06, it can be
seen that the target and calibrator fringe S/N distributions are
almost identical. This suggests that the calibration is
accurate. However, as can be seen, the typical S/Ns are relatively
low; all the distributions peak below 5. Therefore, these observations
are relatively noisy. It can be seen that the observations on
2013-03-04 are of higher quality with typical S/Ns of approximately
10. In this case, the target and calibrator distributions are not
identical but there is considerable overlap, which, in conjunction
with the high S/N, suggests that calibration in this case should be
reliable. The superior S/N of the HD 85567 observations conducted on
2013-03-04 is attributed to the AMBER intervention of January 2013
which improved the sensitivity of AMBER, particularly in the $H-$band
where fringe tracking is conducted.

\smallskip

The disparity between the fringe S/N distributions of the calibrator
and science observations of 2012-05-06 and 2012-12-29 could result in
an inaccurate calibration. This could occur in the following
manner. The lower fringe S/N of the target observations could result
in the target fringes exhibiting a lower fringe contrast than the
calibrators, thus making the target appearing more resolved than the
calibrators, even if this is not the case. Indeed, as noted above, the
raw visibilities of HD 85567 obtained on these two dates are lower
than those of the calibrators. We surmise that the
difference between the target and calibrator observations on these
dates is a bias caused by the fringe tracking performance degrading
when observing the target.

\smallskip

We conclude that only the observations
when the fringe S/N distributions of the target and calibrator
observations are similar can be calibrated. In principle, the
observations of 2012-04-06 and 2013-03-04 offer reliable
calibration. However, since the fringe S/Ns of the observations conducted
2012-04-06 are relatively low, this paper focuses
exclusively on the observations obtained on the date 2013-03-04.

\end{document}